\documentclass[aps,twocolumn]{revtex4}
\usepackage{graphicx,graphics}
\usepackage{amssymb}
\usepackage{amsfonts}
\usepackage{amsmath}

\begin{document}

\title{Longitudinal and transverse static spin fluctuations in layered
ferro- and antiferromagnets }
\author{A. Katanin}
\affiliation{Institute of Metal Physics, 620990, Ekaterinburg, Russia\\
Ural Federal University, 620002, Ekaterinburg, Russia}

\begin{abstract}
We analyse the momentum dependence of static non-uniform susceptibilities of
layered local-moment systems below Curie (Neel) temperature within the $1/S$
expansion, the renormalization-group approach, and first order of $1/N$
expansion. We argue that the previously known results of the spin-wave
theory and renormalization-group approach for the transverse spin
susceptibility acquire strong corrections already at sufficiently low
temperatures, which appear due to the interaction of the incomping magnon
having momentum $q$ with magnons with momenta $k<q.$ Such corrections can
not be treated in the standard renormalization-group approach, but can be
described by both, $1/S$ and $1/N$ expansions. The results of these
expansions can be successfully extrapolated to $T=T_{M},$ yielding the
correct weight of static spin fluctuations, determined by the $O(3)$
symmetry. For the longitudinal susceptibility, the summation of leading
terms of $1/S$ expansion within the parquet approach allows to fulfill the
sumrule for the weights of transverse and longitudinal fluctuations in a
broad temperature region below $T_{M}$ outside the critical regime. We also
discuss the effect of longitudinal spin fluctuations on the (sublattice)
magnetization of layered systems.
\end{abstract}

\maketitle

\bigskip Layered local-moment systems (e.g. layered perovskites\cite{Joungh}
R$_{2}$MO$_{4}$ (R is some element, M is the transition metal) and undoped
parent compounds for high-Tc superconductors) are distinctly different from
the cubic magnets because of the reduced value of magnetic transition (Curie
or Neel) temperature, since it is determined in these systems mainly by weak
interlayer exchange or anisotropy. Relatively small values of magnetic
transition temperatures allow for both theoretical and experimental study of
the evolution of magnetic properties in layered systems in the whole
temperature range $T<T_{N}$ and some temperature region above $T_{N}$.

At low temperatures magnetic excitations in layered (as well, as cubic
systems) are described by the spin wave theory, considering periodic twists
of spins with respect to the ordered state. This theory however does not
describe correctly the thermodynamic properties of layered systems in a
broad temperature range, since it is applicable only at the temperatures,
which are much lower than the magnetic transition temperature. In
particular, the transition temperatures, predicted by this theory appear too
large and the critical exponents are not described correctly (see, e.g.
discussion in Ref. \cite{OurSSWT}).

This situation is reminiscent of weak itinerant magnets\cite{Moriya}, where
the Stoner mean-field theory does not reproduce correctly thermodynamic
properties. For itinerant systems, the shortcomings of Stoner theory had led
to the formulation of spin-fluctuation theory by Murata and Doniach\cite%
{Murata}, Dzyaloshinskii and Kondratenko\cite{Dzyalosh}, and Moriya\cite%
{Moriya}, who considered the effect of collective excitations (paramagnons).
Recently, an improvement of the spin-wave treatment of layered Heisenberg
magnets by considering the effect of spin fluctuations within the RPA
(ladder)-type analysis was proposed in Ref. \cite{OurSSWT}. This analysis
points to a similarity between the two classes of the systems, since it
shows, that (static) longitudinal spin fluctuations in layered Heisenberg
magnets, described by RPA diagrams, play the role, analogous to the
paramagnons in weak itinerant magnets. The corrections to the sublattice
magnetization due to critical (non-spin-wave) fluctuations appear however
also to be important at not too low temperatures, as can be explicitly shown
within the $1/N$ expansion\cite{Our1/N,OurAnis}.

These results motivate to study the effect of different type of excitations
on the momentum $q$-dependence of the static transverse $\chi ^{+-}(\mathbf{q%
},0)$ and longitudinal $\chi ^{zz}(\mathbf{q},0)$ spin susceptibility in
layered systems (second argument corresponds to vanishing frequency). In
particular, while the momentum dependence of $\chi ^{+-}(\mathbf{q},i\omega
_{n})$ is expected to be dominated by one, three, and higher-magnon
processes, $\chi ^{zz}(\mathbf{q},i\omega _{n})$ is determined by two-magnon
processes already in the lowest order of $1/S$ expansion. These two
susceptibilities are not however fully independent, since they must fulfill
the sumrule $\sum_{\mathbf{q},i\omega _{n}}[\chi ^{zz}(\mathbf{q},i\omega
_{n})+\chi ^{+-}(\mathbf{q},i\omega _{n})]=S(S+1),$ which constrains
relative magnitude of longitudinal and transverse spin fluctuations. In
other words, the longitudinal spin fluctuations should be also important for
the transverse susceptibility.

The spin susceptibility of two-dimensional systems was analyzed previously
within the Schwinger boson and modified spin-wave approaches \cite%
{ArovasAuerbach,Yoshioka,Takahashi}, the renormalization-group approach \cite%
{Nelson,CHN} and first-order $1/N$ expansion\cite{Chubukov}. As we discuss
in the present paper, in the symmetric phase the three former approaches
overestimate the spectral weight by a factor $3/2$ (see also Refs. \cite%
{ArovasAuerbach,Yoshioka,KS}), which is important for the fulfillment of the
abovediscussed sumrule. The sumrule appears to be fulfilled in $1/N$
expansion, but violated in the renormalization-group and spin-wave
approaches.

Extending the results for the non-uniform magnetic susceptibilities to the
magnetically-ordered phase allows to study both, the regime of low
temperatures, where the spin-wave approach is applicable, and the critical
regime, where the results of $1/N$ expansion can be applied. Such an
analysis gives also a possibility to investigate the crossover between the
two regimes, which corresponds physically to changing relevant magnetic
fluctuations from spin-wave to the critical ones. The sumrule for spin
susceptibilities can serve as a criterion of the validity of the results of
theoretical approaches.

In the following we analyze the results of $1/S$ expansion for transverse
and longitudinal spin susceptibilities, as well $1/N$ expansion for the
transverse spin susceptibility, which allows us to understand the momentum
dependence of these quantities in different temperature regimes.

\section{The model and the spin-wave approach}

We consider two-dimensional ferro- and antiferromagnets with the easy-axis
anisotropy, described by the Hamiltonian%
\begin{equation}
H=-\frac{1}{2}\sum_{ij}J_{ij}\mathbf{S}_{i}\mathbf{S}_{j}-\frac{1}{2}\eta
J\sum_{ij}S_{i}^{z}S_{j}^{z}-D\sum_{i}(S_{i}^{z})^{2},  \label{H1}
\end{equation}%
where $J_{i,i+\delta _{\bot }}=J,J_{i,i+\delta _{\Vert }}=\alpha J/2$ , and $%
J_{ij}=0$ otherwise$,$ $\delta _{\bot }$ and $\delta _{\Vert }$ being the
vectors, connecting nearest neighbor sites in the same plane and in
different planes; $\eta >0$ and $D>0$ are the two-site and single-site
easy-axis anisotropy parameters.

The $1/S$ expansion can be performed using the Dyson-Maleev representation
\begin{align}
S_{i}^{+}& =\sqrt{2S}a_{i}\,,\;S_{i}^{z}=S-a_{i}^{\dagger }a_{i},
\label{BKJ} \\
S_{i}^{-}& =\sqrt{2S}(a_{i}^{\dagger }-\frac{1}{2S}a_{i}^{\dagger
}a_{i}^{\dagger }a_{i}),  \notag
\end{align}%
where $a_{i}^{\dagger },a_{i}\ $are the Bose ideal magnon operators. To
quadratic order we obtain from Eq. (\ref{H1}) in the ferromagnetic case ($%
J>0 $) the Hamiltonian of interacting spin waves%
\begin{eqnarray}
H &=&H_{\text{sw}}+\frac{1}{4}\sum_{\mathbf{q}_{1}...\mathbf{q}_{4}}\varphi (%
\mathbf{q}_{1},\mathbf{q}_{2}\mathbf{;q}_{3},\mathbf{q}_{4})(a_{\mathbf{q}%
_{1}}^{\dagger }a_{\mathbf{q}_{2}}^{\dagger }a_{\mathbf{q}_{3}}a_{\mathbf{q}%
_{4}})  \notag \\
&&\times \delta _{\mathbf{q}_{1}+\mathbf{q}_{2},\mathbf{q}_{3}+\mathbf{q}%
_{4}},  \label{HInt}
\end{eqnarray}%
where%
\begin{eqnarray}
H_{\text{sw}} &=&\sum_{\mathbf{q}}E_{\mathbf{q}}a_{\mathbf{q}}^{\dagger }a_{%
\mathbf{q}},  \label{HFM} \\
E_{\mathbf{q}} &=&S(J_{0}-J_{\mathbf{q}})+|J|Sf,  \notag
\end{eqnarray}%
$J_{\mathbf{q}}=2J(\cos q_{x}+\cos q_{y})+\alpha J\cos q_{z}$ is the Fourier
transform of the exchange integrals, $f=(2S-1)D/|JS|+(J_{0}/J)\eta S$. The
interaction in Eq. (\ref{HInt}) can be treat perturbatively.

In the case of a two-sublattice antiferromagnet we separate the lattice into
$A$ and $B$ sublattices. On the sublattice $A$ we use the representation (%
\ref{BKJ}), while on the sublattice $B$ the \textquotedblleft
conjugate\textquotedblright\ representation:
\begin{align}
S_{i}^{+}& =\sqrt{2S}b_{i}^{\dagger }\,,\;S_{i}^{z}=-S+b_{i}^{\dagger
}b_{i}^{{}},\;i\in B  \label{BKJb} \\
S_{i}^{-}& =\sqrt{2S}(b_{i}^{{}}-\frac{1}{2S}b_{i}^{\dagger
}b_{i}^{{}}b_{i}^{{}}),  \notag
\end{align}%
where $b_{i}^{\dagger },b_{i}$ are the Bose operators. Introducing operators
$B_{\mathbf{q}}$
\begin{align}
a_{\mathbf{q}}& =(B_{\mathbf{q}}+B_{\mathbf{q+Q}})/2,  \notag \\
b_{-\mathbf{q}}^{\dagger }& =(B_{\mathbf{q}}-B_{\mathbf{q+Q}})/2,
\end{align}%
where $\mathbf{Q}=(\pi ,\pi ,...)$ is the wavevector of the
antiferromagnetic structure, up to a constant term, we have the Hamiltonian
of the same form (\ref{HFM}), but for the operators $B_{\mathbf{q}}$. Note
that in this case $E_{\mathbf{q}}$ in Eq. (\ref{HInt}) does not have a
meaning of an excitation spectrum because of non-Bose commutation relations $%
[B_{\mathbf{q}},B_{\mathbf{p}}^{\dagger }]=\delta _{\mathbf{q,p}}+\delta _{%
\mathbf{q,p+Q}};$ the true excitation spectrum is determined by the
diagonalization of the Hamiltonian with respect to the states with the
momenta $\mathbf{q}$ and $\mathbf{q+Q;}$ in the present paper, however, we
consider only static (classical) contributions to susceptibilities and
(sublattice) magnetization, therefore commutation relations between
operators $B_{\mathbf{q}}$ appear to be not important.

At not too low temperatures ($T\gg T_{q},$ $T_{q}=rJ$ for ferromagnets and $%
T_{q}=r^{1/2}J$ for antiferromagnets, where $r=\max (f,\alpha /2)$) the
result for the sublattice magnetization in the spin-wave theory reads
\begin{equation}
\overline{S}=\langle S_{\mathbf{Q}}^{z}\rangle =\overline{S}_{0}-\frac{T}{%
2\pi JS}\ln \frac{q_{0}}{r^{1/2}},  \label{Sl_magn_SSWT}
\end{equation}%
where $\overline{S}_{0}$ is the ground-state sublattice magnetization, $%
q_{0} $ corresponds to the ultraviolet cutoff parameter in momentum space
for classical spin fluctuations, which is determined by temperature for
quantum magnets ($S\sim 1$): $q_{0}=[T/(JS)]^{1/2}$ \ in the ferromagnetic
quantum case, $q_{0}=T/c$ in the antiferromagnetic quantum case ($c=\sqrt{8}%
|J|S$ is the spin-wave velocity), and $q_{0}=\sqrt{32}$ in the classical
case ($S\gg 1 $), see Ref. \cite{OurSSWT}.

The spin-wave interaction can be treat in the lowest order of $1/S$
perturbation theory the so-called self-consistent spin-wave theory by
performing decoupling of quartic terms in Eq. (\ref{HInt}), see, e.g. Ref.
\cite{OurSSWT}. This theory yields renormalization of the magnon spectrum:
for the renormalized anisotropy and interlayer coupling parameters one finds
$f_{T}=f(\overline{S}/S)^{2},$ $\alpha _{T}=\alpha \overline{S}/S$; these
renormalizations are expected to be qualitatively correct outside the
critical region (see discussion in Ref. \cite{OurSSWT}), in the latter
region the three-dimensional critical fluctuations due to interlayer
coupling or domain wall topological contributions due to easy-axis
anisotropy are expected to be important. For antiferromagnets, one has to
perform additional renormalization $J\rightarrow J\gamma /S$ in the first
term of Eq. (\ref{HFM}), where $\gamma =S+0.079$ is the ground state
(quantum) renormalization of the exchange interaction (its temperature
renormalization can be neglected in the considering cases, in the following
we also assume $\gamma =S$ for ferromagnetic case). In case of considering
the renormalization of the spectrum within the self-consistent spin-wave
theory, the remaining interaction in Eq. (\ref{HInt}) should be considered
as one-particle irreducible, to avoid double counting.

\section{Transverse spin susceptibility}

\subsection{Spin-wave theory}

We consider first the momentum dependence of the static transverse spin
susceptibility $\chi ^{+-}(\mathbf{q},0)=\int_{0}^{1/T}d\tau \langle S_{%
\mathbf{q}}^{+}(\tau )S_{\mathbf{q}}^{-}(0)\rangle .$ Using the
representations (\ref{BKJ}) and (\ref{BKJb}) and decoupling quartic terms
within the spin-wave approach, we obtain for both, ferro- and
antiferromagnets in the small $q$ limit%
\begin{equation}
\chi _{\text{sw}}^{+-}(\mathbf{q},0)=\frac{2\overline{S}}{|J|\gamma q^{2}}
\label{hipm0}
\end{equation}%
(for antiferromagnetic case the shift $\mathbf{q}\rightarrow \mathbf{q}+%
\mathbf{Q}$ is to be performed). Note that the result (\ref{hipm0}) can be
also obtained within the modified spin-wave theory \cite{Takahashi}. The
result (\ref{hipm0}) implies that the static transverse spin susceptibility
in the spin-wave approach would vanish at the magnetic phase transition
temperature, which contradicts the requirement $\chi ^{+-}(\mathbf{q}%
,0)=2\chi ^{zz}(\mathbf{q},0)$ at $T=T_{M}$ and finiteness of the sum of
averaged squares of spin components, see discussion below, in Sect. IIIB.

To overcome the latter drawback of the spin-wave approach, we consider the
lowest-order correction to the result (\ref{hipm0}) due to the magnon
interaction (\ref{HInt})
\begin{eqnarray}
&&\chi _{\text{c-sw}}^{+-}(\mathbf{q},0)%
\begin{array}{c}
=%
\end{array}%
\chi _{\text{sw}}^{+-}(\mathbf{q},0) \\
&&+\frac{T^{2}}{(J\gamma )^{4}}\frac{1}{q^{2}}\sum_{\mathbf{q}_{1}\mathbf{q}%
_{2}}\frac{\varphi (\mathbf{q},\mathbf{q}_{1};\mathbf{q}_{2},\mathbf{q+q}%
_{1}-\mathbf{q}_{2})}{(q_{1}^{2}+r)(q_{2}^{2}+r)((\mathbf{q}+\mathbf{q}_{1}-%
\mathbf{q}_{2})^{2}+r)},  \notag
\end{eqnarray}%
where `c-sw' stands for the spin-wave result, corrected with account of the
spin-wave interaction. For the case of vanishing interlayer coupling we have%
\begin{eqnarray}
\varphi (\mathbf{q}_{1},\mathbf{q}_{2}\mathbf{;q}_{3},\mathbf{q}_{4}) &=&J_{%
\mathbf{q}_{3}}+J_{\mathbf{q}_{4}}-J_{\mathbf{q}_{1}\mathbf{-q}_{3}}-J_{%
\mathbf{q}_{1}\mathbf{-q}_{4}}  \notag \\
&\simeq &-2|J|(\mathbf{q}_{1}\mathbf{q}_{2}+f).  \label{vert}
\end{eqnarray}%
The calculation of the integral yields
\begin{eqnarray}
&&\chi _{\text{c-sw}}^{+-}(\mathbf{q},0)%
\begin{array}{c}
=%
\end{array}
\label{sw1} \\
&&\frac{2}{|J|\gamma q^{2}}\left[ \overline{S}+\frac{1}{\overline{S}_{0}}%
\left( \frac{T}{2\pi |J|\gamma }\ln \frac{q}{f^{1/2}}\right) ^{2}\right] .
\notag
\end{eqnarray}%
This result keeps its form also in the presence of the interlayer coupling
with the replacement $f\rightarrow r.$ We would like to stress that the term
proportional to $\ln ^{2}(q/f^{1/2})$ comes from the integration over
momenta $f^{1/2}<k<q,$ and therefore inaccessible for the standard
renormalization-group techniques\cite{Nelson,CHN}, which consider only
scales, which are larger than all the infrared cutoffs, i.e. $k>\max
(q,f^{1/2}).$

At the same time, as we argue below, the second term in Eq. (\ref{sw1})
appears to be crucially important for fulfillment of the sumrule already by
the lowest-order perturbative correction. Indeed, considering the sum of
squares of the transverse spin components for the result (\ref{sw1}) we
obtain
\begin{align}
& \langle (S_{i}^{x})^{2}\rangle _{\text{stat}}+\langle
(S_{i}^{y})^{2}\rangle _{\text{stat}}%
\begin{array}{c}
=%
\end{array}%
T\sum\limits_{r^{1/2}<q\mathbf{<}q_{0}}\chi _{\text{c-sw}}^{+-}(\mathbf{q,}0)
\notag \\
& =2\overline{S}(\overline{S}_{0}-\overline{S})+\frac{2}{3\overline{S}_{0}}(%
\overline{S}_{0}-\overline{S})^{3}.  \label{SR_lowT}
\end{align}%
The result (\ref{SR_lowT})\ describes correctly not only the low-temperature
behavior of the weight of transverse fluctuations, but also the limit $%
T\rightarrow T_{M}=T_{C}(T_{N}).$ Indeed, in this limit we have $\overline{S}%
=0$ and Eq. (\ref{SR_lowT}) yields $(2/3)\overline{S}_{0}^{2}$ which, as we
see in the following, is the weight, required by the $O(3)$ symmetry.

\subsection{1/N expansion and comparison of different approaches}

To treat the momentum-dependent transverse susceptibility beyond lowest
order of $1/S$-perturbation theory, we perform $1/N$ expansion in a way,
which is similar to that described in Ref. \cite{Chubukov} for 2D
antiferromagnets (see Appendix A). The result of the corresponding expansion
outside the critical regime is%
\begin{equation}
\chi _{1/N,\text{non-crit}}^{+-}(\mathbf{q},0)=\frac{2}{|J|\gamma q^{2}}%
\left[ \overline{S}+\frac{\left( \frac{T}{2\pi \gamma |J|}\ln \frac{q}{%
r^{1/2}}\right) ^{2}}{\overline{S}+\frac{T}{\pi \gamma |J|}\ln \frac{q}{%
r^{1/2}}}\right] .  \label{chi1}
\end{equation}%
It has the same structure, as the result of the $1/S$-perturbation theory (%
\ref{sw1}), except for the denominator in the second term, which accounts
for the effect of longitudinal spin fluctuations, and can be understood in
terms of the vertex corrections, discussed below in Sect. IIIA. On the other
hand, at $T\rightarrow T_{M}$ we obtain
\begin{equation}
\chi _{1/N,T=T_{M}}^{+-}(\mathbf{q},0)=\frac{2}{|J|\gamma q^{2}}\left( \frac{%
aT}{2\pi |J|\gamma }\ln \frac{q}{r^{1/2}}\right) ,~~q>r^{1/2}
\label{chi_ansatz}
\end{equation}%
with $a=2/3$, which is similar to the two-dimensional case\cite{Chubukov}.
The results (\ref{chi1}) at $T\rightarrow T_{M}$ and (\ref{chi_ansatz})
differ only by a coeffitient in front of logarithm: Eq. (\ref{chi1})
corresponds to $a=1/2.$ This difference appears due to the fact, that apart
from the spin-wave and longitudinal excitations, the result of the $1/N$
expansion in the critical regime (\ref{chi_ansatz}) accounts for
non-spin-wave (critical) fluctuations, which change the coefficient in front
of the logarithm to $a=2/3.$

From the result (\ref{chi1}) we find the weight of the transverse spin
fluctuations in $1/N$ expansion outside the critical regime%
\begin{eqnarray}
&&\langle (S_{i}^{x})^{2}\rangle _{\text{stat}}+\langle
(S_{i}^{y})^{2}\rangle _{\text{stat}}%
\begin{array}{c}
=%
\end{array}%
T\sum\limits_{\mathbf{q}}\chi _{1/N}^{+-}(\mathbf{q,}0)  \notag \\
&=&(\overline{S}_{0}^{2}/2)[\sigma +1-2\sigma ^{2}+\sigma ^{2}\arctan \text{h%
}(1-\sigma )],
\end{eqnarray}%
where $\sigma =\overline{S}/\overline{S}_{0},$ while from Eq. (\ref%
{chi_ansatz}) we find at $T\rightarrow T_{N}$
\begin{equation}
\langle (S_{i}^{x})^{2}\rangle _{\text{stat}}+\langle (S_{i}^{y})^{2}\rangle
_{\text{stat}}=T\sum\limits_{\mathbf{q}}\chi _{1/N}^{+-}(\mathbf{q,}0)=a%
\overline{S}_{0}^{2}.
\end{equation}

The weight of the transverse spin fluctuations in the discussed approaches
is shown on Fig. 1. One can see that the weights of the corrected spin-wave
theory and $1/N$ expansion outside the critical regime are numerically close
to each other. Both these theories yield also reasonable extrapolation of
the spectral weight to the limit $T\rightarrow T_{M}$ ($\sigma \rightarrow 0$%
); the momentum dependence of the result of $1/N$ expansion, Eq. (\ref{chi1}%
) is expected to be more correct in the intermediate temperature region,
since it qualitatively reproduces also the limiting momentum dependence, Eq.
(\ref{chi_ansatz}) at $T\rightarrow T_{M}$.
\begin{figure}[tbp]
\includegraphics[width=8.5cm]{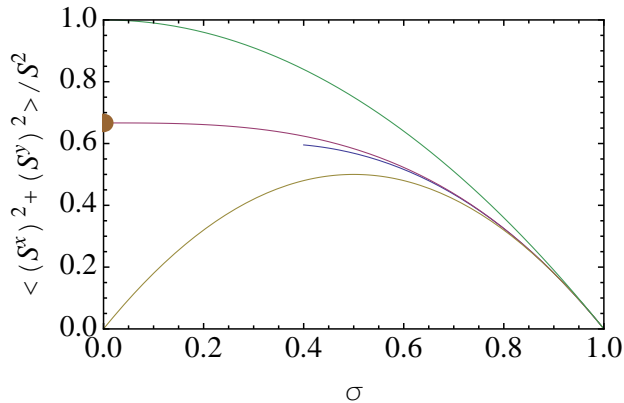}
\caption{(Color online) The weight of the transverse spin fluctuations vs.
(sublattice) magnetisation in different approaches (from top to bottom):
renormalization-group approach, Eq. (\protect\ref{hipmRG}), corrected
spin-wave theory, Eq. (\protect\ref{sw1}), $1/N$-expansion outside the
critical regime, Eq. (\protect\ref{chi1}), and standard spin-wave theory,
Eq. (\protect\ref{hipm0}). The semi-circle at the left axis shows the value
of the weight $a=2/3$, required by O(3) symmetry.}
\end{figure}

The results (\ref{sw1})\ and (\ref{chi_ansatz})\ can be compared to the
result of the renormalization-group analysis\cite{Nelson,CHN,KS}, assuming
that the logarithm in the numerator of Eq. (\ref{hipm0}), contained within
the spin-wave theory in $\overline{S}$, is in fact cut at the largest
low-energy scale, which yields%
\begin{equation}
\chi _{\mathrm{RG}}^{+-}(\mathbf{q},0)=\frac{2}{|J|\gamma q^{2}}\left(
\overline{S}_{0}-\frac{T}{2\pi |J|\gamma }\ln \frac{q_{0}}{q}\right) ,\ \
q>r^{1/2}.  \label{hipmRG}
\end{equation}%
The result (\ref{hipmRG}) can be put in the symmetric phase ($\overline{S}=0$%
) into form (\ref{chi_ansatz}) with $a_{\text{RG}}=1.$ Essentially the same
result was obtained also in the paramagnetic phase within the Schwinger
boson spin-wave approach \cite{ArovasAuerbach,Yoshioka}. (Note that the
modified spin-wave theory \cite{Takahashi} also yields the result (\ref%
{hipmRG}), but for the \textit{longitudinal} spin susceptibility; as
discussed above the transverse spin susceptibility vanishes in the latter
approach in the symmetric phase).

One can see, that the low-temperature results of systematic first-order $1/S$
expansion (\ref{sw1}), as well as $1/N$ expansion, Eq. (\ref{chi1}) are
distinctly different from the results of RG or Schwinger boson approaches,
as they contain second power of logarithm instead of the first. This leads
to a difference of the weight of the transverse spin fluctuations of the
abovementioned approaches already at sufficiently low temperatures ($\sigma $
close to $1$), see Fig. 1. On the other hand, in comparison with the result
of $1/N$ expansion at $T=T_{N}$, Eq. (\ref{chi_ansatz}), the result (\ref%
{hipmRG}) yields different coefficient $a_{\text{RG}}=1$ in Eq. (\ref%
{chi_ansatz}), which yields incorrect weight $\langle (S_{i}^{x})^{2}\rangle
_{\text{stat}}+\langle (S_{i}^{y})^{2}\rangle _{\text{stat}}$ in the RG
approach at $T\rightarrow T_{M}$ \cite{Note}.

\section{Longitudinal spin fluctuations}

\subsection{Spin-wave and RG analysis of longitudinal spin susceptibility}

Let now analyze the longitudinal spin susceptibility within the spin-wave
and RG approaches. The spin-wave theory yields for the spin susceptibility,
longitudinal with respect to the order parameter, the result, which is
expressed as a convolution of the two magnon Green functions. In particular,
in the 2D case with the easy-axis anisotropy we obtain%
\begin{align}
\chi _{0}(q)& =\frac{T}{(J\gamma )^{2}}\sum_{\mathbf{p}}\frac{1}{(\mathbf{p}%
^{2}+f_{T})[(\mathbf{p-q})^{2}+f_{T}]}  \notag \\
\ & \simeq \left\{
\begin{array}{cc}
T/[2\pi (J\gamma q)^{2}]\ln (q^{2}/f_{T}),\; & q^{2}\gg f_{T} \\
\chi _{0}=T/[4\pi (J\gamma )^{2}f_{T}], & q^{2}\ll f_{T}%
\end{array}%
\right. .  \label{Hiq0}
\end{align}%
Similar result with the replacement $f_{T}\rightarrow \alpha _{T}$ can be
also obtained for weak interlayer coupling. In comparison to the transverse
susceptibility, Eq. (\ref{hipm0}), equation (\ref{Hiq0}) has an additional
log-singuliarity in the small momentum limit. As discussed in the previous
Section, they are not sufficient to describe spin fluctuations at not too
low temperatures. In particular, the result (\ref{Hiq0}) does not account
for the interaction of the spin-waves.

To treat the longitudinal spin fluctuations beyond spin-wave theory, the
ladder approximation was suggested\cite{OurSSWT}. This approximation treats
contributions to the vertex of magnon interaction in a certain channel,
corresponding to the scattering of two magnons with close momenta
("longitudinal particle-hole" channel). Below we argue that this
approximation can be rigorously justified by considering logarithmically
large contributions in the parquet-type diagrams, which appear due to
specific momentum dependence of the susceptibility (\ref{Hiq0}).

To perform summation of these contributions, we consider renormalization of
the magnon interaction vertex by magnon scattering in three different
channels. In the lowest order of the perturbation theory we obtain%
\begin{align}
& \Phi (\mathbf{k,p-q;k-q,p})%
\begin{array}{c}
=%
\end{array}%
-2J\mathbf{k}(\mathbf{p-q}) \\
& -4TJ^{2}\sum_{\mathbf{s}}\frac{[\mathbf{k}(\mathbf{s}-\mathbf{q)][s}(%
\mathbf{p}-\mathbf{q)]}}{E_{\mathbf{s}}E_{\mathbf{s-q}}}  \notag \\
& -4TJ^{2}\sum_{\mathbf{s}}\frac{(\mathbf{ks)[(s+k-p)(p-q)]}}{E_{\mathbf{s}%
}E_{\mathbf{s+k-p}}}  \notag \\
& +4TJ^{2}\sum_{\mathbf{s}}\frac{\mathbf{k}(\mathbf{p}-\mathbf{q)}[\mathbf{s}%
(\mathbf{s-k-p+q})\mathbf{]}}{E_{\mathbf{s}}E_{\mathbf{s-k-p+q}}}.  \notag
\end{align}%
Using the relation $\sum_{\mathbf{s}}(s_{a}-q_{a})s_{b}/(E_{\mathbf{s}}E_{%
\mathbf{s-q}})=\delta _{ab}/(4\pi )\ln (q_{0}/q),$ we find to second order
in $1/S$%
\begin{align}
& \Phi (\mathbf{k,p-q;k-q,p})%
\begin{array}{c}
=%
\end{array}%
-2J\mathbf{k}(\mathbf{p-q})  \label{2ndOrder} \\
& \times \left[ 1+\frac{T}{2\pi |J|\gamma ^{2}}\left( \ln \frac{q_{0}}{q}%
+\ln \frac{q_{0}}{|\mathbf{k}-\mathbf{p}|}-\ln \frac{q_{0}}{|\mathbf{k}+%
\mathbf{p-q}|}\right) \right] .  \notag
\end{align}%
To go beyond second order of the $1/S$-perturbation theory, we apply the
one-loop renormalization-group approach, which is also equivalent to
summation of logarithmic divergencies of the parquet diagrams. To this end
we replace infrared cutoffs in the particle-hole and particle-particle
contributions in Eq. (\ref{2ndOrder}) by the cutoff parameter $\mu $,
differentiate r.h.s. of Eq. (\ref{2ndOrder}) over $\mu ,$ and replace the
bare vertices in the right-hand side of equation (\ref{2ndOrder}) by
renormalized vertices [we assume that the momentum dependence
\begin{equation*}
\Phi (\mathbf{k,p-q;k-q,p})=-2J\phi \mathbf{k(p-q)}
\end{equation*}%
does not change, as follows from the lowest-order correction (\ref{2ndOrder}%
)]. In this way we obtain
\begin{equation}
\mu \frac{d\phi }{d\mu }=\frac{T}{2\pi |J|\gamma ^{2}}\phi ^{2}.
\label{RGflow}
\end{equation}%
The equation (\ref{RGflow}) was earlier obtained within the
Holstein-Primakoff representation of spin operators in Ref. \cite{ChubJETP}.
Using of the Dyson-Maleev representation makes however transparent that due
to the change of sign in the particle-particle channel, which occur due to
bilinear structure of the interaction vertex, the contribution of the two of
the three channels cancel each other.

For $q\sim |\mathbf{k}-\mathbf{p}|\sim |\mathbf{k}+\mathbf{p-q}|$ we stop
the flow at $\mu =\max (q,|\mathbf{k}-\mathbf{p}|,|\mathbf{k}+\mathbf{p-q}|)$
to obtain the result%
\begin{equation}
\phi =\frac{1}{1-\frac{T}{2\pi |J|\gamma ^{2}}\ln \frac{q_{0}}{\max (q,|%
\mathbf{k}-\mathbf{p}|,|\mathbf{k}+\mathbf{p-q}|)}}.  \label{RGres}
\end{equation}%
When the momentum transfer in one (some) channel(s) is much smaller than in
the others, one has to continue scaling with the contribution of the
channels with smaller momentum transfer retained. For $q\ll |\mathbf{k}-%
\mathbf{p}|,|\mathbf{k}+\mathbf{p-q}|$ we obtain again the equation (\ref%
{RGres}) with the replacement $\max (q,|\mathbf{k}-\mathbf{p}|,|\mathbf{k}+%
\mathbf{p-q}|)\rightarrow q$. On the other hand, at $q\gg |\mathbf{k}-%
\mathbf{p}|,|\mathbf{k}+\mathbf{p-q}|$ the result (\ref{RGres}) remains
valid, since the contributions of the two remaining channels cancel each
other. Therefore, for arbitrary $q$ and $|\mathbf{k}-\mathbf{p}|\sim |%
\mathbf{k}+\mathbf{p-q}|$ we obtain with the logarithmic accuracy the result
(cf. Ref. \cite{OurSSWT})
\begin{equation}
\Phi (\mathbf{k,p-q;k-q,p})=\frac{|J|\mathbf{k}(\mathbf{q-p})}{\overline{S}%
/(2\gamma )+q^{2}\chi _{0}(q)}+O(|J|f),  \label{Vert_Ladder}
\end{equation}%
where $\overline{S}$ is given by Eq. (\ref{Sl_magn_SSWT}). Thus, as well as
in RPA for itinerant magnets \cite{Moriya}, the effective interaction is
enhanced by fluctuations.

For the static (staggered) non-uniform longitudinal susceptibility (for
antiferromagnetic case the shift $\mathbf{q}\rightarrow \mathbf{q}+\mathbf{Q}
$ has to be performed) we obtain \cite{OurSSWT}
\begin{align}
\chi ^{zz}(q,0)& =\frac{(\overline{S}/\overline{S}_{0})\chi _{0}(q)}{%
1-(T/2\pi |J|\gamma \overline{S}_{0})\ln [q_{0}/\max (\Delta ^{1/2},q)]}
\notag \\
& =\frac{\overline{S}\chi _{0}(q)}{\overline{S}+(|J|\gamma /2)q^{2}\chi
_{0}(q)}.  \label{HiS}
\end{align}%
The vanishing of the weight of longitudinal fluctuations at $T\rightarrow
T_{M}$ is the unphysical result of the parquet approach, which do not
consider critical fluctuations in the vicinity of the magnetic transition
temperature. For the actual layered systems the range $\sigma <0.5$
corresponds to the critical regime, which occur in the rather narrow
temperature range near the transition temperature (the magnetization drops
rather sharply near $T_{\text{M}},$ see Sect. IIIC). The description of this
regime requires considering critical fluctuations of either $d=3$ $O(3)$ or $%
d=2$ $Z_{2}$ symmetry, depending whether interlayer coupling or anisotropy
dominates, see Refs. \cite{OurSSWT,OurRG,Our1/N,OurAnis}. At the same time,
the results (\ref{HiS}) and (\ref{SR2}) are expected to be valid outside the
critical regime, i.e. for $\sigma \gtrsim 0.5$.

\subsection{Sum rule and the weight of longitudinal fluctuations}

In the following we analyze the fulfillment of the sum rule
\begin{eqnarray}
&&\langle (S^{x})^{2}\rangle _{\text{stat}}+\langle (S^{y})^{2}\rangle _{%
\text{stat}}+\langle (S^{z})^{2}\rangle _{\text{stat}}  \label{SR} \\
&=&\sum_{\mathbf{q}}[\chi ^{+-}(\mathbf{q},0)+\chi ^{zz}(\mathbf{q},0)]=%
\overline{S}_{0}^{2}  \notag
\end{eqnarray}%
by the considered approaches; `stat' stands for contribution of classical
spin fluctuations (with the ultraviolet cutoff, given by $q_{0}$). This
sumrule follows from the standard relation $\langle \mathbf{S}^{2}\rangle
=S(S+1);$ the difference $S(S+1)-\overline{S}_{0}^{2}$ in the right-hand
side appears due to retaining only classical term in equation (\ref{SR}).
The possibility of considering separately classical and quantum
contributions is realized for layered systems at temperatures $T_{q}\ll T\ll
|J|S^{2}$ and reflects the `frozenness' of magnetic moments (i.e. separation
of the contributions of quantum and classical spin fluctuations) in the
state with long-range or strong short-range order. The sum rule (\ref{SR})
can be rigorously proven within the nonlinear-sigma model treatment of the
model (\ref{H1}) by excluding the dynamic spin fluctuations which implies
quantum renormalization of the model parameters similarly to the $\phi ^{4}$
model analysis \cite{Sachdev}, see also Refs. \cite{OurRG,KS}.

It can be easily verified that the results of the spin-wave theory (\ref%
{hipm0}) and (\ref{Hiq0}) fulfill the sumrule (\ref{SR}). As discussed in
previous sections, this theory is however expected to be insufficient at not
very low temperatures. Fulfillment of the sum rule (\ref{SR}) by the other
approaches, considered in Sect. II, requires the weight of the longitudinal
fluctuations to be%
\begin{eqnarray}
&&\lbrack \langle (S^{z})^{2}\rangle _{\text{stat}}-\overline{S}^{2}]/%
\overline{S}_{0}^{2}  \label{S^2zz} \\
&=&\left\{
\begin{array}{cc}
(1-\sigma )^{2}(1-\frac{2}{3}(1-\sigma )) & \text{c-sw} \\
\frac{1}{2}[1-\sigma -\sigma ^{2}\arctan \text{h}(1-\sigma )] & \text{%
1/N,non-crit} \\
1/3 & \text{1/N, }T\rightarrow T_{M} \\
0 & \text{RG}%
\end{array}%
\right.  \notag
\end{eqnarray}%
The substruction of $\overline{S}^{2}$ in the left hand side of Eq. (\ref%
{S^2zz}) corresponds extracting the uniform (Bragg) contribution, the result
therefore reflects the contribution of the longitudinal fluctuations. As
discussed above, only the corrected spin-wave theory and the $1/N$ expansion
fulfill the $O(3)$ symmetry, which require that at $T\rightarrow T_{M}$ the
weight of the transverse components should be twice bigge than the
longitudinal ones. For the scaling analysis, the whole spectral weight
(excluding the Bragg contribution) is in fact entirely contained in the
transverse part (see also Ref. \cite{KS}), which explicitly violates $O(3)$
symmentry at $T=T_{N}.$

To verify the conformance of the parquet (RG) approach to longitudinal
fluctuations to the approximations, listed in Eq. (\ref{S^2zz}), we perform
summation over momenta of Eq. (\ref{HiS}) to obtain:
\begin{eqnarray}
T\sum\limits_{\mathbf{q}}\chi _{{}}^{zz}(\mathbf{q,}0) &=&2\overline{S}\left[
\overline{S}_{0}-\overline{S}+\overline{S}\ln (\overline{S}/\overline{S}_{0})%
\right]  \label{SR2} \\
&\rightarrow &\left\{
\begin{tabular}{ll}
$(\overline{S}_{0}-\overline{S})^{2}(1-(2/3)(\overline{S}_{0}-\overline{S}%
)),\ $ & $\overline{S}\rightarrow \overline{S}_{0}$ \\
$2\overline{S}_{0}\overline{S}$ & $\overline{S}\rightarrow 0$%
\end{tabular}%
\right. .  \notag
\end{eqnarray}%
Comparing the first row of the second line of Eq. (\ref{SR2}) with Eq. (\ref%
{S^2zz}) we find that at low temperatures it agrees with the corrected
spin-wave analysis of the transverse part implying that the considered
parquet approach yields correct asympthotic behavior of the spectral weight
at low temperatures. In fact, the corresponding range of sublattice
magnetizations $\sigma >0.4,$ where the results for the weight of
longitudinal fluctuations in the parquet analysis, corrected spin-wave
theory and $1/N$ expansion outside the critical regime agree with each
other, appears to be rather broad (see Fig. 2); as discussed above, it
corresponds to the temperatures below the (narrow) critical regime near $%
T_{M}$. As discussed in Sect. IIa, the result of the low-temperature $1/N$
expansion (\ref{chi1}) also allows to fulfill approximately the sumrule at $%
\sigma >0.4;$ while the result (\ref{chi_ansatz}) fulfills the sum rule at $%
T=T_{M}$.
\begin{figure}[tbp]
\includegraphics[width=8.5cm]{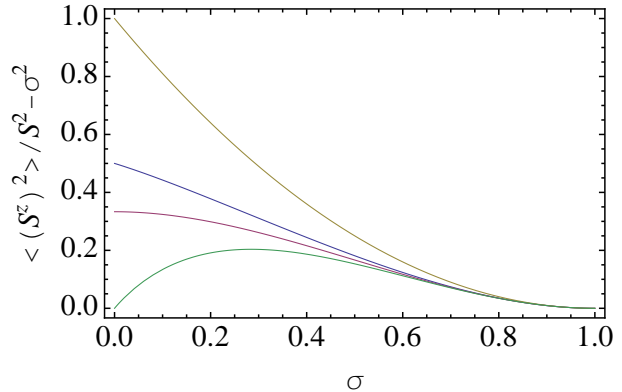}
\caption{(Color online) The weight of the longitudinal fluctuations vs.
(sublattice) magnetisation. Lines from top to bottom correspond to: standard
spin-wave theory, Eq. (\protect\ref{Hiq0}), applying the sumrule (\protect\ref{SR}) to the results
of the $1/N$-expansion outside the critical regime and corrected spin-wave
theory (see Eq. (\protect\ref{S^2zz})), and the weight obtained in the
parquet (RG) approach, Eq. (\protect\ref{HiS}). }
\end{figure}

\subsection{The effect on the (sublattice) magnetization}

Finally, we discuss the effect of longitudinal static spin fluctuations on
the (sublattice) magnetization. It was argued in Ref. \cite{OurSSWT} that
the longitudinal spin fluctuations yield the subleading logarithmic
correction to the spin-wave result (\ref{Sl_magn_SSWT}), which improves
comparison of the result with the experimental data\cite{Chubukov1}. Such a
correction can be also naturally obtained in the two-loop
renormalization-group approach\cite{OurRG}, which yields
\begin{eqnarray}
\sigma ^{1/\beta _{2}} &=&1-\frac{T}{4\pi \rho _{s}}\left[ (N-2)\ln \frac{%
q_{0}^{2}}{\Delta (T)}\right.  \notag \\
&&+2\ln \frac{1}{\max (\sigma ^{1/\beta _{2}},T/(4\pi \rho _{s}))}
\label{Res1} \\
&&\left. -2(1-\sigma ^{1/\beta _{2}})+F\left( T/(4\pi \rho _{s}\sigma
^{1/\beta _{2}})\right) \right] ,  \notag
\end{eqnarray}%
where $\rho _{s}$ is the ground-state spin stiffness (e.g. $\rho
_{s}=J\gamma \overline{S}_{0}$ in the self-consistent spin-wave theory), $%
\Delta (T)=\max (\alpha _{T}/2,f_{T}),$ and $F\left( x\right) $ is some
non-singular function, which accounts for the contribution of higher-loop
terms. The \textquotedblleft critical exponent\textquotedblright\ $\beta
_{2},$ which is the limit of the exponent $\beta _{2+\varepsilon }$ in $%
d=2+\varepsilon $ dimensions at $\varepsilon \rightarrow 0,$ is given by $%
\beta _{2}=(N-1)/(2(N-2)).$ At $N=3$ we have $\beta _{2}=1$. The leading
logarithmic term in Eq. (\ref{Res1}) corresponds in this case to the
self-consistent spin-wave theory, while the subleading logarithmic term
coincides with the result of the parquet approximation, discussed in Sect.
IIIB (cf. Ref. \cite{OurSSWT}), and therefore describes the contribution of
the static longitudinal fluctuations to (sublattice) magnetization. The
nonsingular term $-2(1-\sigma )$ does not follow directly from the parquet
approach and at low temperatures compensates the second term in the square
brackets. The function $F$ stands for the contribution of non-parquet
diagrams, not accounted by renormalization-group analysis.

The result (\ref{Res1}) can be compared to that of the first-order $1/N$
expansion of the sublattice magnetization, which reads\cite{Our1/N,OurAnis}
\begin{align}
& \sigma =\left\{ 1-\frac{T}{4\pi \rho _{s}}\left[ (N-2)\ln \frac{2T^{2}}{%
\Delta (\alpha _{r},f_{r})}+B_{2}\ln \frac{1}{\sigma ^{2}}\right. \right.
\label{magn_first_1/N} \\
& \left. \left. -2(1-\sigma ^{2})-I_{1}(x_{\sigma })\right] \right\} ^{\beta
_{2}},  \notag
\end{align}%
where $\alpha _{r}=\alpha _{T=0}$ and $f_{r}=f_{T=0}$ are the ground-state
quantum-renormalized parameters of the anisotropy and interlayer coupling, $%
B_{2}=3+f_{r}/\sqrt{f_{r}^{2}+2{\alpha }_{r}f_{r}},$
\begin{equation}
x_{\sigma }=\frac{4\pi \rho _{s}}{(N-2)T}\sigma ^{2}.
\end{equation}%
For the cases, when only anisotropy or interlayer coupling is present ($%
\alpha _{r}=0$ or $f_{r}=0$), the results (\ref{Res1}) and (\ref%
{magn_first_1/N}) differ by the replacement $\sigma ^{1/\beta
_{2}}\rightarrow \sigma ^{2}$ in the right-hand side, which is related to
the property of the first-order $1/N$ expansion, that the $N=\infty $ value
of the exponent $\beta _{2}=1/2$ is corrected only in the leading
(proportional to $N-2$), but not in the subleading (proportional to $N^{0}$)
terms in Eq. (\ref{magn_first_1/N}). Although for the abovementioned cases
the result (\ref{Res1}) can be therefore considered as a self-consistent
modification of the first-order $1/N$ result, it appears that
non-self-consistent equation (\ref{magn_first_1/N}) better agree with
experimental data (see below), which can be related to the importance of
non-spin-wave fluctuations, not captured by the renormalization-group
approach.

Let us also discuss the relation of the results of 1/$N$ expansion for
sublattice magnetization in the presence of interlayer coupling only ($%
f_{r}=0$) to the approach, proposed recently in Ref. \cite{Sushkov}. The
result of Ref. \cite{Sushkov} up to the logarithmic accuracy reads
\begin{equation}
\sigma =\left( 1-\frac{T}{2\pi \rho _{s}}\ln \frac{2T^{2}}{c^{2}\alpha
\sigma }\right) ^{1/2}.  \label{magn_r=2_Sushkov}
\end{equation}

\begin{figure}[tbp]
\includegraphics[width=8cm]{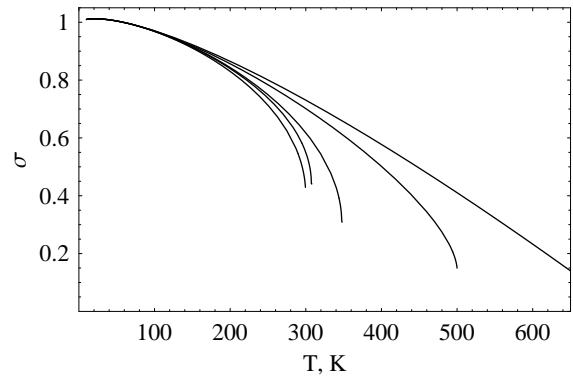}
\caption{Temperature dependence of sublattice magnetization of La$_2$CuO$_4$
in different approaches (from top to bottom): spin-wave theory, Eq. (
\protect\ref{Sl_magn_SSWT}), self-consistent spin-wave theory, Eq. (
\protect\ref{Sl_magn_SSWT}) with the replacement $\protect\alpha \rightarrow
\protect\alpha _T$, renormalization-group approach, Eq. (\protect\ref{Res1}%
), 1/N expansion, Eq. (\protect\ref{magn_first_1/N}), and the approach of
Ref. \protect\cite{Sushkov}, Eq. (\protect\ref{magn_r=2_Sushkov}) of the
present paper}
\label{Fig4}
\end{figure}

This result can be compared to the results of first-order $1/N$ expansion (%
\ref{magn_first_1/N}). One can see, that although Eq. (\ref{magn_r=2_Sushkov}%
) contains some of the first-order $1/N$ corrections, it borrows the
exponent of magnetization $1/2$ from the result of the zeroth-order $1/N$
expansion in the intermediate temperature regime and neglects subleading
terms in Eq. (\ref{magn_first_1/N}), which account for the longitudinal spin
fluctuations.

Comparison of different approaches is presented in Fig. 3. We choose the
parameters, which correspond the compound La$_{2}$CuO$_{4}$: $\rho _{s}=$%
290K, $c=$2618K, and $\alpha =10^{-3}$. One can see consequent reduction of
the Neel temperature from spin-wave and self-econsistent spin-wave analysis,
and further to the renormalization-group approach. The results (\ref%
{magn_first_1/N}) and (\ref{magn_r=2_Sushkov}) appear to be numerically
close to each other, and therefore Eq. (\ref{magn_r=2_Sushkov}) provides
successful extrapolation of the sublattice magnetization from its
low-temperature limit.

\section{Conclusion}

In the present paper we have considered momentum dependence of static
transverse- and longitudinal spin susceptibilities, analyzed the fulfillment
of the sumrule (\ref{SR}) and discussed the effect of longitudinal spin
fluctuations on the (sublattice) magnetization.

For the transverse susceptibility, we have shown that the result of the
corrected spin-wave theory, accounting for the spin-wave interaction in the
lowest order in $1/S$, yields the correct weight of transverse spin fluctuations
in both, the low-temperature limit ant $T\rightarrow T_M$.
The correction to the pure spin-wave result is proportional to $\ln
^{2}(q/r^{1/2})$ and comes from the integration over the range $r^{1/2}<k<q$
of the momenta $k$ of virtual magnons, interacting with the physical state
at momentum $q.$ This momentum range is inaccessible for the standard
renormalization-group approach, which treats only contribution of momenta $%
k>q.$ Therefore, standard
renormalization-group approach by Nelson and Pelkovitz \cite{Nelson} and
Chakraverty, Halperin, and Nelson \cite{CHN}, applied to the non-uniform spin susceptibility,
has to be supplemented by considering the RG flow at
the scales $k<q.$ This can be done in particular within the functional
renormalization-group analysis\cite{Wetterich}, which is leaved as a subject
for future studies.

The first-order $1/N$ expansion allows to determine the static transverse
spin susceptibility outside the critical regime and in the limit of Curie
(Neel) temperature. The corresponding correction to the spin-wave result for
the transverse nonuniform spin susceptibility changes from $\ln
^{2}(q/r^{1/2})$ at low temperatures to $\ln (q/r^{1/2})$ near $T=T_{M}$.
The result of $1/N$ expansion at $T=T_{M}$ differs by a factor $2/3$ from
the results of Schwinger-boson and the renormalization-group analysis, which
happens because the latter approaches assume spin-wave picture of the
excitation spectrum instead of a critical one. Further analysis of the
crossover between the spin-wave and critical regime seems to be of the
certain interest.

The applicability of the ladder (parquet) approximation for the longitudinal
spin susceptibility was analyzed. It was shown that the results of these
approximations allow to fulfill the sumrule at the temperatures outside the
critical regime, where the relative (sublattice) magnetization $\sigma >0.4$%
. Extending the results of $1/N$ analysis to the order $1/N^{2}$, which
would make possible to describe both, longitudinal and transverse spin
susceptibilities within this approach with higher accuracy, are of the
certain interest. The extension of the presented analysis to calculate the
\textit{dynamic} magnetic susceptibilities on the basis of the discussed
approaches would allow to study the difference of the spin-wave and critical
dynamics.

Acknowledgements. The author is grateful to V. Yu. Irkhin, A. F. Barabanov,
A. N. Ignatenko, and O. Sushkov for stimulating discussions. The work is
supported by the Partnership program of the Max-Planck Society.

\section*{Appendix A. 1/N expansion for the transverse spin susceptibility}

Following to \cite{ArovasBook,Chubukov,Our1/N}, to construct $1/N$ expansion
we pass to the continuum classical $O(N)$-model with the partition function
\begin{eqnarray}
Z &=&\int D\sigma D\lambda \exp \left\{ -\frac{\rho _{s}}{2}%
\int\limits_{0}^{1/T}d\tau \int d^{2}\mathbf{r}\right.  \label{zp} \\
&&\ \ \ \ \ \ \left. \left[ (\mathbf{\nabla }\sigma )^{2}+f(\sigma
_{1}^{2}+..+\sigma _{N-1}^{2})+i\lambda (\sigma ^{2}-1)\right] \right\} ,
\notag
\end{eqnarray}%
where $\sigma =\{\sigma _{1}...\sigma _{N}\}$ is the vector field, $\rho
_{s} $ is the quantum-renormalized spin stiffness, $f$ is the dimensionless
anisotropy parameter (see main text, we consider here for simplicity the
two-dimensional case $\alpha =0$). In the ordered phase (only this will be
considered) the excitation spectrum in the zeroth order in $1/N$, which is
given by the poles of the unperturbed longitudinal and transverse Green's
functions, contains a gap $f^{1/2}$ for all the components $\sigma _{m}$
except for $m=N$:
\begin{equation}
G_{t}^{0}(q)=\left( q^{2}+f\right) ^{-1},\,\,\,G_{l}^{0}(q)=q^{-2}.
\end{equation}

The (dimensionless) self-energy of the transverse fluctuations is given by
\begin{equation}
\Sigma _{t}(k)=\frac{2T}{N}\int \frac{d^{2}\mathbf{q}}{(2\pi )^{2}}\frac{%
G_{t}^{0}(\mathbf{k-q})}{\widetilde{\Pi }(q)},  \label{Sigma}
\end{equation}%
where
\begin{equation}
\widetilde{\Pi }(q)=T\int \frac{d^{2}\mathbf{p}}{(2\pi )^{2}}%
G_{t}^{0}(p)G_{t}^{0}(\mathbf{q}+\mathbf{p})+\frac{2\sigma ^{2}}{\widetilde{g%
}}G_{l}^{0}(q),  \notag
\end{equation}%
and $\widetilde{g}=(N-1)/\rho _{s}$ is the quantum-renormalized coupling
constant. Note that Eq. (\ref{Sigma}) represents a contribution of
longitudinal spin fluctuations to the transverse spin susceptibility, as $%
\widetilde{\Pi }^{-1}(q)$ has a very similar structure to the vertex (\ref%
{Vert_Ladder}), renormalized by the longitudinal fluctuations.

Evaluation of the integral in (\ref{Sigma}) yields%
\begin{eqnarray}
\Sigma _{t}(k) &=&-\frac{k^{2}}{N}\ln \left( \frac{T\widetilde{g}}{2\pi }\ln
\frac{k}{f^{1/2}}+\sigma ^{2}\right)  \notag \\
&&+\frac{k^{2}}{N}\frac{\ln (k/f^{1/2})}{\ln (k/f^{1/2})+2\pi \sigma ^{2}/(T%
\widetilde{g})}  \notag \\
&&+\frac{k^{2}}{N}O(\frac{1}{\ln (k/f^{1/2})+2\pi \sigma ^{2}/(T\widetilde{g}%
)}).  \label{Sigma1}
\end{eqnarray}%
Using the result of the zeroth-order $1/N$ expansion
\begin{equation}
\sigma ^{2}=1-(\widetilde{g}T/(2\pi ))\ln (q_{0}/f^{1/2})  \label{sigma}
\end{equation}%
(which is consistent with the calculating the first-order $1/N$ correction
to the self-energy), we represent Eq. (\ref{Sigma1}) in the form%
\begin{eqnarray}
&&\Sigma _{t}(k)%
\begin{array}{c}
=%
\end{array}%
-\frac{k^{2}}{N}\ln \left( 1-\frac{T\widetilde{g}}{2\pi }\ln \frac{q_{0}}{k}%
\right)  \label{SigmaRes} \\
&&+\frac{k^{2}}{N}\left( \frac{\ln (k/f^{1/2})}{\ln (k/f^{1/2})+2\pi \sigma
^{2}/(T\widetilde{g})}\right) +\frac{k^{2}}{N}O(\frac{1}{\ln (k/f^{1/2})}).
\notag
\end{eqnarray}%
The first term in Eq. (\ref{SigmaRes}) can be transformed to the form, given
by the renormalization-group approach (generalization of the Eq. (\ref%
{hipmRG}) to arbitrary $N$); to this end we perform in Eq. (\ref{SigmaRes})
the standard replacement $N\rightarrow N-2$ \cite{Chubukov,Our1/N}, which is
required because of considering first-order $1/N$ terms only, and collect
logarithm into the power. Collecting other terms to comply the lowest orders
of perturbation result (\ref{sw1}) yields%
\begin{eqnarray}
\chi ^{+-}(\mathbf{q},0) &=&\frac{N-1}{\rho _{s}q^{2}}\left( 1-\frac{T%
\widetilde{g}}{2\pi }\ln \frac{q_{0}}{q}\right) ^{1/(N-2)}  \notag \\
&&\times \left[ 1+\frac{T\widetilde{g}}{2\pi }\frac{\ln (q/f^{1/2})}{\sigma
^{2}+\frac{T\widetilde{g}}{2\pi }\ln \frac{q}{f^{1/2}}}\right] ^{-1/(N-2)}.
\notag \\
&=&\frac{N-1}{\rho _{s}q^{2}}\left[ \sigma ^{2}+\frac{\left( \frac{T%
\widetilde{g}}{2\pi }\ln \frac{q}{f^{1/2}}\right) ^{2}}{\sigma ^{2}+\frac{T%
\widetilde{g}}{\pi }\ln \frac{q}{f^{1/2}}}\right] ^{1/(N-2)}  \label{1NlowT}
\end{eqnarray}%
Replacement $N\rightarrow N-2$ and collecting terms into power in the
multiplier in the second line of Eq. (\ref{1NlowT}) is however unjustified
near the magnetic transition temperature, since in that case this multiplier
describes contribution of non-spin-wave degrees of freedom, which reduces
the transverse susceptibility $(N-1)/N$ times\cite{Chubukov}. Instead, we
find in that case, similarly to the two-dimensional case\cite{Chubukov}%
\begin{eqnarray}
\chi ^{+-}(\mathbf{q},0) &=&\frac{N-1}{\rho _{s}q^{2}}  \label{chi_1/N} \\
&&\times \left[ \frac{N-1}{N}\left( \frac{T\widetilde{g}}{2\pi }\ln \frac{q}{%
f^{1/2}}\right) \right] .  \notag
\end{eqnarray}%
Since the aboveconsidered calculation accounted for logarithmic terms only,
the same results with the replacement $f\rightarrow r$ hold in the presence
of the interlayer coupling. Comparing this to the spin-wave results (\ref%
{hipm0}) and (\ref{sw1}) one should have in mind that $N=3$ in the physical
case and the used zeroth-order $1/N$ result for $\sigma ^{2}$ (\ref{sigma})
is analogous to the spin-wave result for $\overline{S}/\overline{S}_{0}$.


\begin{thebibliography}{99}
\bibitem{Note} Note that it was incorrectly concluded in Ref. \cite{Chubukov}%
, that the result of the RG approach agrees with the result of the $1/N$
expansion, Eq. (\ref{chi_ansatz}). In fact, the result of Ref. \cite{CHN}
(Eqs. (D11) and (D12)) differ by a factor $(N-1)/N$ from that of Eqs. (5.21)
and (5.22) in Ref. \cite{Chubukov} ($N=3$ for the Heisenberg model), while
both papers \cite{CHN,Chubukov} considered the correlation function of a
single spin component.
\end{thebibliography}
\end{document}